\documentclass[12pt]{scrartcl}
\usepackage{color}
\usepackage{url}
\usepackage{amssymb}
\usepackage{amstext}
\usepackage{relsize}
\usepackage{graphicx}
\def \asy#1#2{^{+#1}_{-#2}}
\newcommand{\stat}{{\rm (stat)}}
\newcommand{\syst}{{\rm (syst)}}
\newcommand{\invf}{{\rm fb}^{-1}}

\newcommand{\Bstar}{B^{*}}
\newcommand{\Bbar}{\bar{B}}

\newcommand{\Bs}{B_{s}}

\newcommand{\Bsbar}{\bar{\Bs}}

\newcommand{\mumu}{\mu^+\mu^-}

\def \etalbelle{{\it et\,al.} [Belle Collaboration]}
\def \etalcleo{{\it et\,al.} [CLEO Collaboration]}
\def \etalbabar{{\it et\,al.} [BaBar Collaboration]}

\def \etald0{{\it et\,al.} [D0 Collaboration]}

\def \etale835{{\it et\,al.} [Fermilab E835 Collaboration]}

\def\babar{\mbox{\slshape B\kern-0.1em{\smaller A}\kern-0.1em B\kern-0.1em{\smaller A\kern-0.2em R}}}

\newcommand{\gev}{\,\mathrm{ GeV}}

\newcommand{\gevm}{\mathrm{ GeV}/c^2}

\newcommand{\ee}{e^+e^-}

\newcommand{\hb}{h_b({\mathrm{1P}})}

\newcommand{\hbp}{h_b({\mathrm{2P}})}
\newcommand{\hbn}{h_b({\mathrm{mP}})}

\newcommand{\chib}{\chi_b({\mathrm{1P}})}
\newcommand{\chibp}{\chi_b({\mathrm{2P}})}
\newcommand{\etab}{\eta_b({\mathrm{1S}})}
\newcommand{\etabp}{\eta_b({\mathrm{2S}})}
\newcommand{\etabm}{\eta_b({\mathrm{m^\prime S}})}

\newcommand{\pipm}{\pi^{\pm}}

\newcommand{\goesto}{\rightarrow}

\newcommand{\jpsi}{\mbox{$ J/\psi$}}

\newcommand{\piz}{\mbox{$\pi$}^{0}}
\newcommand{\pipi}{\mbox{$\pi$}^{+}\mbox{$\pi$}^{-}}
\newcommand{\pizpiz}{\mbox{$\pi$}^{0}\mbox{$\pi$}^{0}}
\newcommand{\upsid}{\mbox{$\Upsilon$}{\rm (1D)}}

\newcommand{\upsi}{\mbox{$\Upsilon$}{\rm (1S)}}
\newcommand{\upsii}{\mbox{$\Upsilon$}{\rm (2S)}}
\newcommand{\upsiii}{\mbox{$\Upsilon$}{\rm (3S)}}
\newcommand{\upsiv}{\mbox{$\Upsilon$}{\rm (4S)}}
\newcommand{\upsv}{\mbox{$\Upsilon$}{\rm (5S)}}
\newcommand{\upsns}{\mbox{$\Upsilon$}{\rm (nS)}}

\newcommand{\chibj}{\mbox{$\chi_{bJ}({\mathrm{1P}})$}}

\newcommand{\mmpp}{\mbox{$M_{\rm miss}$}}
\newcommand{\mmissn}{\mbox{$\mmpp^{(m)}$}}

\newcommand{\bea}{\begin{eqnarray}}
\newcommand{\beq}{\begin{equation}}
\newcommand{\eea}{\end{eqnarray}}
\newcommand{\eeq}{\end{equation}}

\newcommand{\upsot}{\Upsilon({\mathrm{1S,2S}})}

\newcommand{\upsott}{\Upsilon({\mathrm{1S,2S,3S}})}

\begin{document}

\title{RESULTS IN $\mathbf{B_s}$ PHYSICS AND \\ BOTTOMONIUM SPECTROSCOPY \\ USING THE BELLE $\mathbf{\Upsilon(5S)}$ DATA}

\author{
CHRISTIAN OSWALD\thanks{oswald@physik.uni-bonn.de, Physikalisches Institut, Universit\"{a}t Bonn, Bonn, D-53115, Germany.}
\and
TODD K. PEDLAR\thanks{todd.pedlar@luther.edu, Department of Physics, Luther College, Decorah, IA, 52101, USA.}
}

\maketitle

\begin{abstract}
The Belle experiment at KEK accumulated a $121.4~\invf$ sample of $\ee$ collisions
at the $\Upsilon(5S)$ resonance. This sample provides ample opportunity for improving
the understanding of both the properties of $\Bs$ mesons and the spectroscopy of bottomonium states.
In this article we describe the recent results obtained from the Belle $\Upsilon(5S)$ data. 
\end{abstract}

\newpage

\section{Introduction}

The story of $\Upsilon(5S)$ measurements began in 1985 when
the CLEO and CUSB collaborations reported the first ``observation of a new structure in the $e^+e^-$ cross
section above the $\Upsilon(4S)$'', with $0.1~{\rm fb}^{-1}$ of data collected at the Cornell Electron
Storage Ring.\cite{Besson:1984bd,Lovelock:1985nb}
The observed $\Upsilon(5S)$ resonance is a bottomonium state (quark content $b\bar{b}$) with a
mass of $(10876 \pm 11)~{\rm MeV}/c^2$ and a width of $(55 \pm 28)~{\rm MeV}/c^2$.\cite{pdg}
In the ensuing two decades, CLEO and the $e^+e^-$ $B$-factory experiments, Belle and \babar~at
KEKB (Tsukuba, Japan) and PEPII (Stanford, US), focussed on taking data at energies
near the mass of the $\Upsilon(4S)$ resonance, which is just above the kinematic threshold for pair production of $B^0\bar{B}^0$ and $B^+B^-$ pairs and therefore provides an ideal environment to study the decays of these mesons. The heavier $B_s$ mesons cannot
be produced at $\Upsilon(4S)$, but it is possible at higher collision energies. It was in 2003 when a larger
$\Upsilon(5S)$ data sample of $0.42~{\rm fb}^{-1}$ was collected by the CLEO III detector.
Analysis of this data led to the first evidence for $B_s$ production at the $\Upsilon(5S)$ energy.\cite{artuso2005,Bonvicini:2005ci}
This sparked interest in the $\Upsilon(5S)$ resonance at the Belle experiment and the first $23.6~{\rm fb}^{-1}$ data sample was collected from 2005 to 2006 and then extended to a total integrated luminosity of $121.4~{\rm fb}^{-1}$ collected from 2008 to 2009. Belle's $\Upsilon(5S)$ data sample is by far the largest to date,
and all results in this review are obtained from it.

While the main goal of the $\Upsilon(5S)$ physics program was to study the decays of the
$B_s$ meson, it turned out that Belle's large data sample is also an ideal base to study
the bottomonium spectrum and to discover unexpected states. We structure this review
in two parts: in the first part, we discuss the
$B_s$ measurements, and in the second, the spectroscopy of conventional
and unconventional bottomonium states.

Before presenting the results, we clarify some of the terminology used throughout this
review. What is commonly considered as the $\Upsilon(5S)$ peak in the $e^+e^-$ hadronic cross section
might contain a contribution from a neighbouring $Y_b$ state which is discussed in Sec. \ref{sec:yb}.
Besides the resonant process $e^+e^- \to \Upsilon(5S)$ and possibly $e^+e^- \to Y_b$,
the hadronic cross section contains the so-called continuum
processes $e^+e^- \to q\bar{q}$, with $q = u,d,s,c,b$. In the literature the label ``$\Upsilon(5S)$'' is used
for the whole bottom production near the $\Upsilon(5S)$ mass peak,
including all resonant and non-resonant processes -- in other words for everything but
the $q\bar{q}$ continuum involving quarks lighter than the $b$-quark ($q=u,d,s,c$).
We will follow this convention in this article.

\section{$B_s$ Measurements}

The $B$-factory experiments Belle and \babar~can look back on a veritable success
story. Their measurements of $B^0$ and $B^+$ meson decays
significantly improved our understanding of $CP$ violation and
quark flavor transitions described by the Cabibbo-Kobayashi-Maskawa (CKM) mechanism.\cite{CKM1,CKM2}
The knowledge of $B_s$ decays was, however, rather sparse until recently when
Belle started pioneering the investigation of $B_s$ mesons using the $\Upsilon(5S)$
data sample. Shortly after, the experiments at the Large Hadron Collider (LHC) located near Geneva
followed, in particular the dedicated $B$-physics experiment LHCb.
The LHC experiments profit from an enormously high $B_s$ production cross section,
$\sigma(pp \to B_s X) = (0.105 \pm 0.013)\times 10^5~{\rm nb}$, at $\sqrt{s}=7~{\rm TeV}$ in
$pp$-collisions,\cite{Aaij:2013noa} compared to $\sigma(e^+e^- \to B_s X) = (0.12 \pm 0.02)~{\rm nb}$ 
at the $\Upsilon(5S)$ resonance.\cite{BsCrosss} However,
studying $B_s$ decays at an $e^+e^-$ $B$-factory has certain advantages over measurements
at a hadron collider. Firstly, the number of produced
$b$-flavored mesons is well known and absolute branching fractions can be
measured. In contrast to the production at hadron colliders, $B_s$ mesons are produced
in $\Upsilon(5S)$ decays coherently in quantum mechanically entangled pairs.
Full or partial reconstruction of one of the $B_s$ mesons provides information on the second
$B_s$ meson in the event, which is a prerequisite for inclusive analyses,
for example the measurement of the $B_s \to X \ell \nu$ branching fraction discussed in Sec. \ref{sec:bs2xlnu}.
A further benefit of an $e^+e^-$ collider is the complete knowledge
of the initial state that provides kinematic constraints for the reconstruction of
undetected particles such as neutrinos.

\begin{figure}[tb]
\centering
\includegraphics[width=5.0in]{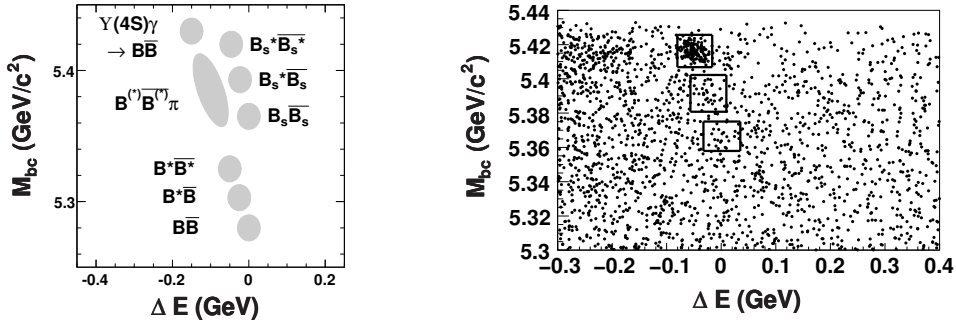}
\caption{The sketch on the left visualises the preferred location of fully reconstructed
$B_{(s)}$ candidates in the $\Delta E$-$M_\text{bc}$ plane for the different
$\Upsilon(5S)$ decay modes. The 2D distribution on the right shows
$B_s \to D_s^-\pi^+$ candidates in the $\Delta E$-$M_\text{bc}$ plane
with the $B_s^*\bar{B}_s^*$, $B_s^*\bar{B}_s$ and $B_s\bar{B}_s$ signal regions indicated by boxes.
The $B_s$ candidates were reconstructed in $23.6~{\rm fb}^{-1}$ of
$\Upsilon(5S)$ data collected with the Belle detector.\protect\cite{Louvot:2008sc}\label{fig:DeltaE_Mbc}
Figure courtesy of the Belle Collaboration.}
\end{figure}

The analyses of $B_s$ decays with $\Upsilon(5S)$ data build on the experience of $B^0$ and $B^+$ studies at $\Upsilon(4S)$
by transferring the existing techniques to 
the higher collision energy. Most $B_s$ analyses performed so far are untagged, {\it i.e.}
one $B_s$ is fully reconstructed in the signal final state, while the second $B_s$ in the event
is not explicitly reconstructed. Correctly reconstructed $B_s$ mesons can be separated from
misreconstructed candidates by means of two variables:
the beam energy constrained mass, $M_\text{bc} = \sqrt{s/4 - p_B^{*2}}$, and the
difference between expected and reconstructed $B_s$ energy,
$\Delta E = E_B^* - \sqrt{s/2}$. The variables $p_B^*$ and $E_B^*$ are the
reconstructed $B_s$ momentum and energy in the center-of-mass frame of the colliding beams, respectively.
In $\Upsilon(4S)$ decays, the $B\bar{B}$ pairs are produced close to the kinematic threshold,
and decay nearly at rest in the $e^+e^-$ rest frame. Correctly reconstructed $B$ candidates have thus
$M_\text{bc}$ around the nominal $B$ mass and $\Delta E$ consistent with zero.
The situation is a little different in $\Upsilon(5S)$ decays, where not only the production of
$B_s\bar{B}_s$ pairs is kinematically allowed, but also $B_s^*\bar{B}_s$ and
$B_s^* \bar{B}_s^*$. The mass difference between $B_s^*$ and $B_s$ is only $48.7^{+2.3}_{-2.1} {\rm MeV}/c^2$,\cite{pdg}
thus the photon emitted in the $B_s^*$ decay has too low energy to be efficiently reconstructed. Hence, true $B_s$ candidates
populate three distinct regions in the $\Delta E$-$M_\text{bc}$ plane depending on whether the $\Upsilon(5S)$ decay was
to $B_s\bar{B}_s$, $B_s^*\bar{B}_s$ or $B_s^*\bar{B}_s^*$ (see Fig. \ref{fig:DeltaE_Mbc}).

\subsection{Estimation of $B_s$ production}

The $\Upsilon(5S)$ resonance decays to $B_s^{(*)}\bar{B}_s^{(*)}$ pairs
as well as to $B^{(*)}\bar{B}^{(*)}$ pairs and to final states with bottomonia, discussed
later in this paper: the relative production fractions are denoted by $f_s$, $f_{ud}$
and $f_{\not{B}}$, respectively, and by definition they sum up to unity: $f_s + f_{ud} + f_{\not{B}} = 1$.
The parameter $f_s$ is a key ingredient to
calculate the total $B_s$ yield in the sample, $N(B_s) = 2 \cdot \mathcal{L} \cdot \sigma_{b\bar{b}} \cdot f_s$,
where $\mathcal{L}$ is the integrated luminosity of the data and $\sigma_{b\bar{b}}$ is the cross section
of the process $e^+e^- \to b\bar{b}$. The untagged branching fraction measurements 
presented in this review require $N(B_s)$ as normalization and thus rely on $f_s$. Since the
parameter $f_s$ plays such a central role for the normalization of the measurements, we discuss
its determination in the following.

The principle of all $f_s$ measurements is to compare decay rates to a chosen final state measured in a $B_s$-enriched and
a $B_s$-depleted sample. One measures for example the inclusive $D_s$ rates in data
recorded at $\Upsilon(5S)$ ($B_s$~enriched) and $\Upsilon(4S)$ ($B_s$~depleted).\cite{artuso2005,Drutskoy:2006fg}
The value of $f_s$ can be extracted from such measurements using the relation:
\begin{equation}
\mathcal{B}(\Upsilon(5S) \to D_s X) =
2 \cdot f_s \cdot \mathcal{B}(B_s \to D_s X)
+ f_{ud} \cdot \mathcal{B}(\Upsilon(4S) \to D_s X))\,.
\end{equation}
The production fraction of $B^0$ and $B^+$ mesons at $\Upsilon(5S)$,
$f_{ud}$, can be replaced by $(1 - f_s - f_{\not{B}})$, where
$f_{\not{B}}$ is estimated from the sum of measured branching fractions to bottomonium states.\cite{Amhis:2012bh}
Belle applied this method to the full $121.4~{\rm fb}^{-1}$ data sample collected at the
$\Upsilon(5S)$ resonance and obtained the value $f_s = (17.2~\pm~3.0)\%$, corresponding to $(7 \pm 1)$ million $B_s^{(*)}\bar{B}_s^{(*)}$ pairs.
The dominant uncertainty in this kind of $f_s$ measurement arises from the uncertainty on the prediction for the
branching fraction $\mathcal{B}(B_s \to D_s X)$, which is taken from
a model-dependent estimate.\cite{artuso2005,suzuki1984}
There are variants of this $f_s$ measurement that use inclusive $D^0$, $\phi$ or $B$
yields.\cite{Drutskoy:2006fg,Huang:2006em,Drutskoy:2010an} The available published measurements
were combined and result in $f_s = (19.9 \pm 3.0)\%$.\cite{pdg}

An approach that avoids the dependence on hadronic branching fractions, which are difficult to predict, 
uses dilepton events where the leptons stem from semileptonic $B_{(s)}$ decays.\cite{Sia:2006cq,phdlouvot}
The inclusive semileptonic $B_s$ branching fraction can be estimated with relatively high
precision from the well measured $B^0$ branching fraction, and the $B^0$ and $B_s$
lifetimes $\tau_s$ and $\tau_d$ assuming SU(3) flavor symmetry:
\begin{equation}
\mathcal{B} (B_s \to X \ell \nu) =
\mathcal{B} (B^0 \to X \ell \nu) \cdot \frac{\tau_s}{\tau_d}\,.
\label{eq:bsinclrel}
\end{equation}
The charge of the lepton ($\ell = e,~\mu$) from the semileptonic decay is sensitive to the flavor of
the decaying $b$-quark. Since the mixing probability for a $B_s$ meson, $\chi_s = (49.9309 \pm 0.0012)\%$,
is much higher than for a $B^0$ meson, $\chi_d = (18.75 \pm 0.20)\%$,\cite{pdg}
and no mixing occurs for $B^+$ mesons, the measured rates of
dilepton events with same-sign $\ell^\pm\ell^\pm$ pairs ($B_s$-enhanced) and opposite-sign
$\ell^\pm\ell^\mp$ pairs ($B_s$-depleted) can be used to extract $f_s$. So far,
no results with this method have been published. The expected precision on $f_s$ is
$10$ to $15\%$, which would be equal to or better than the combination of all existing measurements.

Not only the fraction of events containing $B_s^{(*)}$ mesons, but also the
fractions of the different modes $B_s\bar{B}_s$, $B_s^*\bar{B}_s$ and $B_s^*\bar{B}_s^*$
are of interest. These can be obtained from a fit to the $\Delta E$-$M_\text{bc}$ distribution
of fully reconstructed $B_s \to D_s^- \pi^+$ decays (see Fig. \ref{fig:DeltaE_Mbc}).\cite{Louvot:2008sc}
The most common mode is $B_s^*\bar{B}_s^*$ with a fraction of $f_{B_s^*\bar{B}_s^*}=(87.0 \pm 1.7)\%$.\cite{fBsBs}

\subsection{$CP$ violating decay modes}

Decay time distributions of $B_s$ decays to final states such as $J/\psi \phi$
give access to the $CP$ violating phase of $B_s^0\bar{B}_s^0$ oscillations, $\phi_s$,
and the decay width difference $\Delta \Gamma_s = \Gamma_L - \Gamma_H$
of the light and heavy $B_s$ mass eigenstates. Measurements of such $B_s$ decay time
distributions are the domain of the LHCb experiment as it collects large $B_s$
samples with high boosts of the $B_s$ mesons, and has an excellent timing resolution sensitive
to the fast $B_s$-oscillations.
Measurements with $\Upsilon(5S)$ data can contribute to a better understanding of the 
resonance structure in the decays and provide measurements of the absolute branching fractions. The measurements in this section are untagged and use the variables $M_\text{bc}$ and $\Delta E$ for signal extraction.

The $J/\psi \phi$ final state is a superposition of $CP$-even and $CP$-odd states
and therefore $\phi_s$ and $\Delta \Gamma_s$ have to be extracted in an angular analysis.\cite{Aaij:2013oba}
Precise knowledge of the underlying resonant and non-resonant backgrounds is vital in
this procedure. Belle studied the contribution of $\phi \to K^+K^-$,
$f_2^\prime(1525) \to K^+K^-$ and the remaining non-resonant components
to the total $B_s \to J/\psi K^+K^-$ decay width by fitting the different contributions
to the $K^+K^-$ invariant mass distribution.\cite{Thorne:JpsiKK}
The underlying S-wave component in the $\phi$ mass window is measured
taking into account the possibility of an additional $B_s \to J/\psi f_0(980)$ component.
This approach provides complementary information to the time dependent angular
analysis by LHCb.\cite{Aaij:2013orb}

\begin{figure}[tb]
\centering
\includegraphics[width=5.0in]{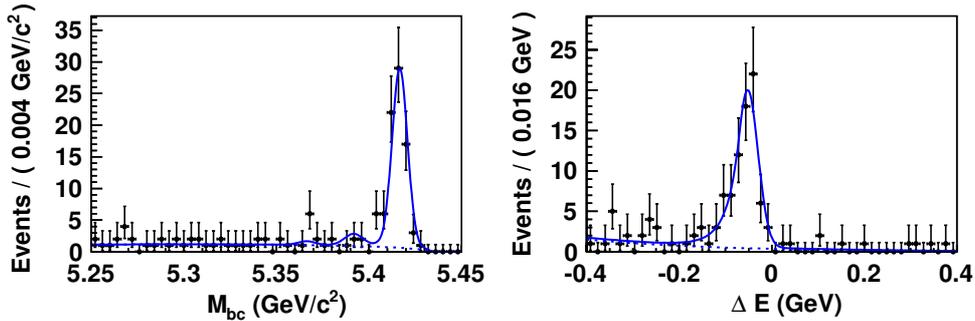}
\caption{$M_{\rm bc}$ and $\Delta E$ distribution of $B_s \to J/\psi \eta (\gamma \gamma)$ events
reconstructe in $121.4~{\rm fb}^{-1}$ of $\Upsilon(5S)$ data collected with the Belle detector.
The distributions show projections to the $B_s^*\bar{B_s}^*$ signal region of the other
variable, $\Delta E \in (-116; 12)$ MeV and $M_{\rm bc} \in (5.405; 5.428)$ GeV$/c^2$.
The solid curves are the projections of the fit result, the dotted curves represent the background
shape. \label{fig:jpsieta}
Reprinted figure with permission from J.~Li {\it et al.}  [Belle Collaboration],
Phys. Rev. Lett. 108, 181808, 2012. Copyright (2012) by the American Physical Society.}
\end{figure}

An angular analysis is not necessary to determine the $CP$ violating parameters, if the $B_s$ decays to
a $CP$ eigenstate, for example $J/\psi f_0(980)$. The $B_s \to J/\psi f_0(980)$ decay 
was observed in 2011 at the same time
by LHCb and Belle.\cite{Aaij:2011fx,Li:2011pg} Belle also claimed evidence for the decay $B_s \to J/\psi f_0(1370)$.\cite{Li:2011pg}
Moreover, Belle reported the first observation of the $B_s$ decays to the $CP$-even states
$J/\psi \eta$, with $\eta \to \gamma \gamma$ and $\eta \to \pi^+\pi^-\pi^0$ (see Fig. \ref{fig:jpsieta}), and
$J/\psi \eta^\prime$, with $\eta^\prime \to \eta \pi^+ \pi^-$ and $\eta^\prime \to \rho^0 \gamma$.\cite{Belle:2012aa}
A further study was dedicated to the decays of $B_s$ mesons to
$K^+K^-$ and $K^0\bar{K}^0$.\cite{Peng:2010ze}

The $\Upsilon(5S)$ data sample also allowed for the measurement of the $B_s \to D_s^{(*)+}D_s^{(*)-}$
branching fractions.\cite{Esen:2012yz,Esen:2010jq} Such decay modes can be measured with less model
dependence than at hadron colliders, due to the presence of two low-energy photons.
The measurements put constraints on the parameter space of the decay width difference $\Delta \Gamma_s$ and the
angle $\phi_{12}=\arg(M_{12}/\Gamma_{12})$, where $M_{12}$ and $\Gamma_{12}$ are the off-diagonal
elements of the $B_s$ mass and decay matrices. Under certain theory assumptions it can be deduced
from this measurement that $\phi_{12} \lesssim 40^\circ$.\cite{Chua:2011er}

The measured branching fractions of the $CP$ violating $B_s$ decay modes are summarized in Table
\ref{tab:cpviol}. It is worthwhile pointing out that the systematic uncertainties of all measurements
are dominated by the $\sim15\%$ uncertainty on $f_s$.
Current methods of $f_s$ determination are statistically limited and substantial
improvement of the precision can be expected with a larger $\Upsilon(5S)$ data sample
at a next generation $B$-factory.

\begin{table}[tb]
\centering
\caption{Branching fraction measurements of $CP$ violating decay modes. All results are obtained using the full Belle $\Upsilon(5S)$ data sample ($121.4~{\rm fb}^{-1}$). The statistical and systematic uncertainties of the branching fractions are added quadratically.}
\begin{tabular}{lccc}
\hline \hline
$B_s$ decay mode & Branching fraction $[10^{-3}]$ & Signal yield & Ref.\\
\hline
$J/\psi \eta$ &
$\hphantom{0}0.51\hphantom{0}~{}^{+}_{-}~\hphantom{0}{}^{0.13\hphantom{0}}_{0.10\hphantom{0}}$  &
$141 {}^{+}_{-} 14$ &
\cite{Belle:2012aa}\\

$J/\psi \eta^\prime$ &
$\hphantom{0}0.37\hphantom{0}~{}^{+}_{-}~\hphantom{0}{}^{0.10\hphantom{0}}_{0.09\hphantom{0}}$ &
$\hphantom{0}86 {}^{+}_{-} 14$  &
\cite{Belle:2012aa}\\

$J/\psi f_0(980),~f_0(980) \to \pi^+\pi^-$ &
$\hphantom{0}0.12\hphantom{0}~{}^{+}_{-}~\hphantom{0}{}^{0.04\hphantom{0}}_{0.03\hphantom{0}}$ &
$\hphantom{0}63 {}^{+16}_{-10}$ &
\cite{Li:2011pg} \\

$J/\psi f_0(1370),~f_0(1370) \to \pi^+\pi^-$ &
$\hphantom{0}0.034~{}^{+}_{-}~\hphantom{0}{}^{0.014}_{0.015}$ &
$\hphantom{0}19 {}^{+6\hphantom{0}}_{-8\hphantom{0}}$ &
\cite{Li:2011pg}\\

$J/\psi \phi$ &
$\hphantom{0}1.25\hphantom{0} ~{}^{+}_{-}~\hphantom{0}0.24\hphantom{0}$ &
$326 {}^{+}_{-} 19$ &
\cite{Thorne:JpsiKK} \\

$J/\psi f_2^\prime(1525)$ &
$\hphantom{0}0.26\hphantom{0}~{}^{+}_{-}~\hphantom{0}0.08\hphantom{0}$ &
$\hphantom{0}60 {}^{+}_{-} 13$ &
\cite{Thorne:JpsiKK} \\

$J/\psi K^+K^-$ &
$\hphantom{0}1.01\hphantom{0}~{}^{+}_{-}~\hphantom{0}0.22\hphantom{0}$ &
$536 {}^{+}_{-} 32$ &
\cite{Thorne:JpsiKK} \\

$K^+K^-$ &
$\hphantom{0}0.038~{}^{+}_{-}~\hphantom{0}{0.012}$ &
$\hphantom{0}23 {}^{+}_{-}6$ &
\cite{Peng:2010ze}\\

$K^0\bar{K}^0$ &
$< \hphantom{0}0.066$ at $90\%$ C.L.&
$\hphantom{0}\hphantom{0}5 {}^{+5}_{-4}$ &
\cite{Peng:2010ze}\\

$D_s^+D_s^-$ &
$\hphantom{0}5.8\hphantom{0}\hphantom{0}~{}^{+}_{-}~\hphantom{0}{}^{1.7\hphantom{0}\hphantom{0}}_{1.6\hphantom{0}\hphantom{0}}$ &
$\hphantom{0}33 {}^{+6}_{-5}$ &
\cite{Esen:2012yz} \\

$D_s^{*\pm}D_s^{\mp}$ &
$18\hphantom{.}\hphantom{0}\hphantom{0}\hphantom{0}~{}^{+}_{-}~\hphantom{0}5\hphantom{.}\hphantom{0}\hphantom{0}\hphantom{0}$ &
$\hphantom{0}45 {}^{+}_{-} 6$ &
\cite{Esen:2012yz} \\

$D_s^{*+}D_s^{*-}$ & $20\hphantom{.}\hphantom{0}\hphantom{0}\hphantom{0}~{}^{+}_{-}~\hphantom{0}6\hphantom{.}\hphantom{0}\hphantom{0}\hphantom{0}$ & $\hphantom{0}24 {}^{+}_{-} 4$ &
\cite{Esen:2012yz} \\

$D_s^{(*)+}D_s^{(*)-}$ &
$43\hphantom{.}\hphantom{0}\hphantom{0}\hphantom{0}~{}^{+}_{-}~11\hphantom{.}\hphantom{0}\hphantom{0}\hphantom{0}$ &
---
&
\cite{Esen:2012yz} \\
\hline \hline
\end{tabular}
\label{tab:cpviol}
\end{table}

\subsection{Semileptonic decays}\label{sec:bs2xlnu}
\begin{figure}[tb]
\centering
\includegraphics[width=4.0in]{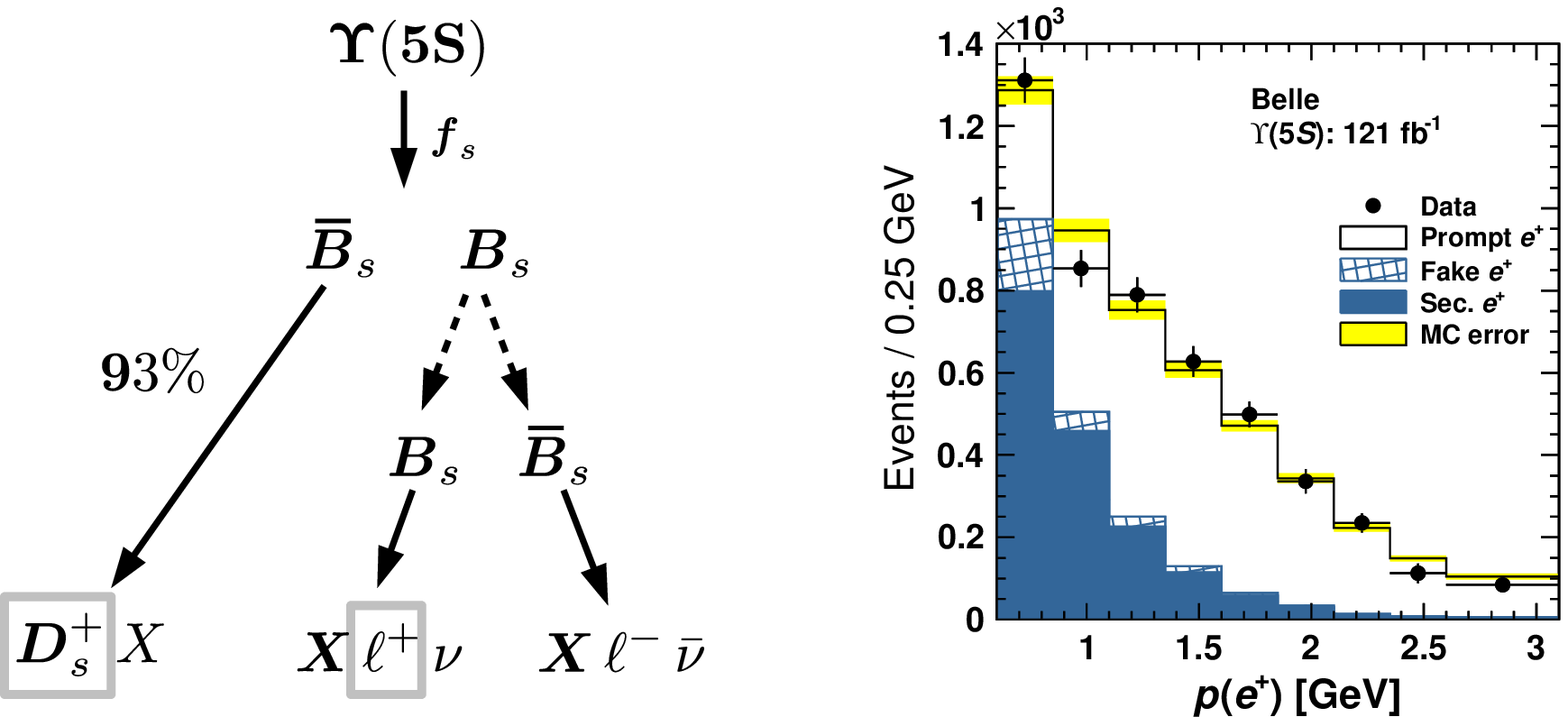}
\vspace*{8pt}
\caption{Measurement of the $B_s \to X \ell \nu$ branching fraction: The diagram
on the left illustrates how the selection of same-sign $D_s^+\ell^+$ combinations
ensures that the signal lepton $\ell^+$ and the tag $D_s^+$ meson stem from different
$B_s$ decays. The distribution on the right
shows the measured electron momentum spectrum for events with correctly reconstructed
tag $D_s^+$. \label{fig:semilep}
Reprinted figure with permission from C.~Oswald {\it et al.}  [Belle Collaboration],
Phys. Rev. D 87, 072008, 2013. Copyright (2013) by the American Physical Society.}
\end{figure}

Semileptonic decays of $B_{(s)}$ mesons are a powerful tool to determine the
elements $|V_{cb}|$ and $|V_{ub}|$ of the CKM matrix,
to probe the quark dynamics inside the $B_{(s)}$ meson
and to study $CP$ violation. The inclusive branching
fraction, $\mathcal{B}(B_s \to X \ell^+ \nu)$, where $X$ is an arbitrary hadronic final state
and $\ell=e,~\mu$, is an important parameter in the determination
of the $B_s$ production fraction at the $B$-factories and the LHC.\cite{Aaij:2011jp}
The SU(3) symmetry relation given by Eq. \ref{eq:bsinclrel} is often used in these measurements
to estimate the branching fraction $\mathcal{B}(B_s \to X \ell^+ \nu)$. Theory calculations predict
that this equality holds at the percent level,\cite{Bigi:2011gf,Gronau:2010if} but this has to be
proven in experimental tests. The \babar~collaboration measured the branching fraction
$\mathcal{B} (B_s \to \ell \nu X) = (9.5 {}^{+2.5}_{-2.0}(\text{stat}) {}^{+1.1}_{-1.9}(\text{syst}))\%$
with $4.25~{\rm fb}^{-1}$ of data collected in the center-of-mass energy range
between $10.54~{\rm GeV}$ and $11.20~{\rm GeV}$.\cite{Lees:2011ji}

The Belle collaboration profited from their large $\Upsilon(5S)$ data sample to perform the most
precise measurement of the $B_s \to X \ell \nu$ branching fraction.\cite{Oswald:2012yx}
This decay mode has a large event yield due to the large expected branching fraction
($\sim 10\%$ for each flavor, $e$ and $\mu$), and as only the charged lepton is reconstructed, it can be
detected with high efficiency.
Consequently, it was feasible to tag $B_s\bar{B}_s$ pair events
by reconstruction of $D_s^+$ mesons from the Cabibbo-favored decay mode $\bar{B}_s \to D_s^\pm X$,
which has a large probability of $(93 \pm 25)\%$.\cite{pdg} The $D_s^+$ tag enhances the relative number of
$B_s$ mesons in the sample from $20\%$ to approximately $70\%$.
The tag $D_s^+$ mesons are reconstructed in the clean $D_s^+ \to \phi(\to K^+K^-) \pi^+$ decay mode.
To ensure that the tag $D_s^+$ and the signal lepton $\ell^+$ stem from different
$B_s$ mesons, they are required to have the same sign of the electric charge,
as illustrated in Fig.~\ref{fig:semilep} (left). The same-sign requirement implies that due to
$B_s$ mixing, only $\chi_s \approx 50\%$ of the signal leptons are selected.

Two samples were analysed: one containing all $D_s^+$ candidates and the
other all $D_s^+\ell^+$ candidates. The yield of correctly reconstructed $D_s^+$
mesons is obtained from fits to the $K^+K^-\pi^+$ invariant mass distributions.
The $D_s^+\ell^+$ sample contains not only signal leptons, but also secondary leptons
from decays of $B_{(s)}$ daughters and misidentified lepton candidates. The yield of signal leptons
is obtained from a fit to the lepton momentum spectrum (see Fig~\ref{fig:semilep} (right)).

The $B_s \to X \ell \nu$ branching fraction is calculated from the efficiency-corrected
$D_s^+$ and $D_s^+\ell^+$ yields. The yields include contributions from the $B^0$ and $B^+$ decays:
$B~\to~D_s^\pm~X$ and $B~\to~X~\ell~\nu$. The fraction of $B_s$ events in the
$D_s^+$ and $D_s^+\ell^+$ samples is estimated from external measurements, including $f_s$ and
the $B_s$ mixing probability. The external parameters
and the resulting uncertainties are listed in Table \ref{tab:bs2xlnupars}.
The extracted $B_s~\to~X~\ell~\nu$ branching fraction is $(10.6 \pm 0.5(\text{stat}) \pm 0.7(\text{syst}))\%$,
in agreement with the theory predictions.\cite{Bigi:2011gf,Gronau:2010if}

This analysis is a good example for the benefits of the coherent production of
$B_s^{(*)}\bar{B}_s^{(*)}$ pairs at an $e^+e^-$ $B$ factory.
Tagging one $B_s$ meson in the event with a reconstructed $D_s^+$ meson allows one not
only to study semileptonic $B_s$ decays inclusively,
it also reduces the systematic uncertainty on the $B_s$ production
fraction, which is $\sim 6 \%$, compared to $\sim 15\%$ in the untagged measurements.

\begin{table}[tb]
\centering
\caption{Parameters for the extraction of the $B_s \to X \ell \nu$ branching fraction and the resulting relative uncertainties.}
\begin{tabular}{lc}
\hline \hline
Parameter(s) & Relative uncertainty on $\mathcal{B} (B_s \to X \ell \nu)$ \\
\hline 
$B_s$ production fraction: $f_s$ &  $2.4\%$ \\
$B^0$, $B^+$ production fraction: $f_u$, $f_d$ & $1.0\%$ \\
$B_s \to D_s^\pm X$ multiplicity & $4.4\%$ \\
$B \to D_s X$ branching fractions & $3.1\%$ \\
$B \to X \ell \nu$ branching fractions & $0.4\%$\\
$B^{(*)}\bar{B}^{(*)}(\pi)$ hadronization fractions & $0.3\%$ \\
$B^0$ and $B_s$ mixing probabilities & $0.2\%$ \\
\hline \hline
\end{tabular}
\label{tab:bs2xlnupars}
\end{table}

\subsection{Hadronic decays}

Studies of the hadronic decays $B_s \to D_s^{(*)} h$ with $h = \pi^+,~K^\pm,~\rho^+$ were amongst
the earliest $B_s$ measurements performed with $23.6~{\rm{fb^{-1}}}$ of the Belle $\Upsilon(5S)$ data set and
will not be discussed here.\cite{Louvot:2008sc,Louvot:2010rd}
These measurements test theory calculations that predict similar branching fractions
for $B_s$ and $B^0$ decays based on SU(3) symmetry.\cite{Deandra:1993}

There have also been studies of baryonic $B_s$ decay modes.
Baryonic decays of lighter $b$-flavored mesons, for example $B^+ \to \Lambda_c^- p \pi^+$, were discovered before,
and an enhancement of the branching fraction in the
baryon-antibaryon mass spectrum near the kinematic threshold was observed.\cite{Dytman:2002yd,Abe:2004sr,Aubert:2008ax}
This aroused interest in studies of the corresponding $B_s$ decays. Recently, Belle found evidence for
the decay $B_s \to \Lambda_c^- \Lambda \pi^+$ and measured the branching fraction
$(3.6 \pm 1.1(\text{stat}) \pm 1.2(\text{syst})) \times 10^{-4}$.\cite{Solovieva:2013rhq}
Unfortunately, the $\Upsilon(5S)$ data sample is not large enough to make
a statement on the phenomenon of a branching fraction enhancement near threshold
as observed for $B$ decays.

\subsection{Radiative penguin decays}

Measurements of branching fractions and kinematic spectra of processes that are
suppressed in the Standard Model (SM)
are a promising indicator for ``new physics'' because the decay rate could be
significantly enhanced by loop contributions involving particles beyond
the SM. In $B_s$ decays, unique $b \to s$  flavor changing neutral
current transitions can be studied, such as $B_s \to \phi \gamma$ and $B_s \to \gamma \gamma$.
In the SM, these decays proceed via so-called penguin diagrams, whose branching
fractions were predicted to be at the order of $4 \times 10^{-5}$ and $5 \times 10^{-7}$, respectively.\cite{Ali:2007sj,Reina:1997my}
In the initial $23.6~{\rm fb}^{-1}$ data sample collected at $\Upsilon(5S)$, Belle observed the decay
$B_s \to \phi \gamma$ for the first time and measured its branching fraction to be $(57 {}^{+18}_{-15}(\text{stat}) {}^{+12}_{-11}(\text{syst})) \times 10^{-6}$,
consistent with the SM expectation, which was later confirmed by LHCb.\cite{Wicht:2007ni,LHCb:2012ab}
No significant signal was observed for the $B_s \to \gamma\gamma$ decay and the upper limit of $8.7 \times 10^{-6}$ on the branching fraction was set at the $90\%$ confidence level.\cite{Wicht:2007ni}

\section{Bottomonium Spectroscopy at $\upsv$\label{sec:spec} }

A data sample obtained from $\ee$ collisions at the mass of the $\upsv$ would not have been
selected {\em a priori} as a likely sample with which to pursue spectroscopy of narrow bottomonia.
Both the $\upsiv$ and $\upsv$, being above open-flavor threshold, dominantly decay to open bottom
mesons: $B^{(*)}\bar B^{(*)},\Bs\Bsbar$.  The partial widths of $\upsiv$ and $\upsv$ to $\pipi\upsns$ (where n = 1, 2, 3)
were expected to be similar to those for $\upsiii$ to $\pipi(\upsii,\upsi)$ - which implies an expected branching
fraction for such transitions that would be far too small for lower bottomonium states to be profitably studied
through them.  In studies both by \babar\cite{Aubert08a} and Belle\cite{Sokolov09}, the
partial widths for $\upsiv\to\pipi(\upsii,\upsi)$ were measured and satisfied these expectations.  However, the expectations
were shown to be completely incorrect in the case of $\upsv$.  In the following sections
we describe both the measurement of these anomalously large $\upsv\to\pipi(\upsiii,\upsii,\upsi)$ transition rates in the Belle data
as well as the subsequent discovery of several conventional and unconventional bottomonium states and transitions among
them.

\subsection{Observation of anomalously large $\pipi$ transition rates from $\upsv$ to lower bottomonia\label{sec:pipirates}}

Using a sample of $21.7~\invf$ of $\ee$ collisions at the $\upsv$,
Belle observed very large rates for the
transitions $\upsv\to\pipi\upsott$: up to 100 times larger than
the corresponding rates for $\pipi$ transitions between
$\upsns$ (n = 1, 2, 3) states that lay below the open-bottom threshold,
or even from $\upsiv$.\cite{Chen08}   Immediately upon this
observation, speculation ensued concerning the explanation of this wildly unexpected
result.  Several explanations were offered, including
the existence of tetraquark or other exotic non-$\upsns$ states near
$\upsv$.\cite{Ali10}

\begin{figure}[tb]
\centering
\includegraphics[width=0.6\textwidth]{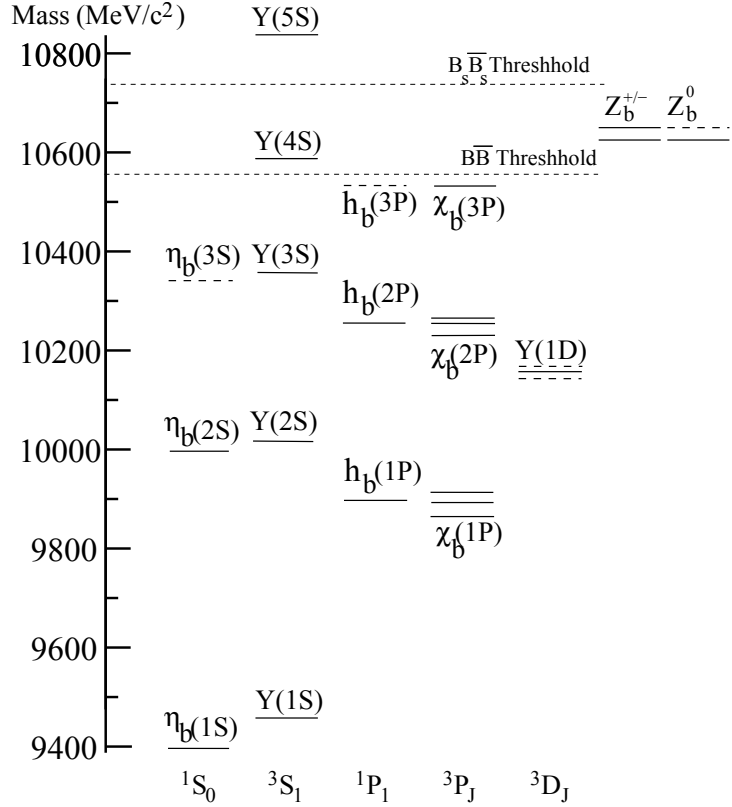}
\vspace*{8pt}
\caption{Current status of bottomonium and bottomonium-like states at or below the mass of $\Upsilon(5S)$
(dashed states not yet observed).
The states are arranged as columns of like spin and orbital angular momentum, and the text beneath each column
indicates these quantum numbers, with the usual spectroscopic notation, $^{2S+1}L_{J}$, where $S$, $L$ and $J$ are the
total spin, orbital and total angular momentum, respectively.  To indicate the value of $L$, the usual letters
S, P and D indicate states of orbital angular momentum 0, 1 and 2, respectively.  
\label{fig:ups}}
\end{figure}

\subsection{Spectroscopy of conventional bottomonium states\label{sec:convspec}}

The currently-known bottomonium spectrum is shown in Fig.~\ref{fig:ups}.  The S-wave triplet states, $\upsns$, and
the P-wave triplet states, $\chib$ and $\chibp$, have been known for many years, while the ground state S-wave singlet, $\etab$,
was discovered in $\Upsilon(3S) \to \etab \gamma$ decays only in 2008 by \babar~and confirmed by
CLEO.\cite{babaretab,babaretab2,cleoetab}  The bottomonium states recently discovered using the $\upsv$ data sample at Belle
are the two lower-lying P-wave singlets, $\hb$ and $\hbp$, and the second S-wave singlet, $\etabp$.  The discoveries of
these states are detailed in the following sections.

\subsubsection{Discovery of singlet-$P$ states $\hbn$\label{sec:singlet-p}}

The observation of large rates for $\pipi$ transitions of $\upsv$ to lower vector bottomonia 
made attractive the possibility of searching for transitions to other, previously unobserved 
states such as the singlet-P and excited singlet-S states. Additional motivation for these 
searches came from the CLEO observation of the production of the
singlet-P charmonium state $h_c$ via the process $e^+ e^-
\to \pi^+ \pi^- h_c$ at a center-of-mass energy of 4.16 GeV, lying
above charm threshold.\cite{Pedlar11}   The rate for this transition is comparable to that for $e^+ e^- \to \pi^+
\pi^- J/\psi$, which is surprising, since the transition to $h_c$ requires a constituent charm quark to undergo a
spin flip and should therefore be suppressed relative to the transitions between vector charmonia, which do not
require a charm quark spin flip.

The search for $\hbn$ states (where m = 1, 2) was done using a hadronic event selection in which at least one oppositely-charged
pair of positively-identified pions was observed.   Only the information from the two charged pions was used, and 
the yield of $\hbn$ production was obtained from the spectrum of the $\pipi$ missing mass, which is
defined as
\begin{equation}
M_\text{miss} \equiv \sqrt{[P(\upsv) - P(\mathrm{\pipi})]^2} = \sqrt{[M(\upsv)-E^*_{\pipi}]^2-[p^*_{\pipi}]^2},\\
\end{equation}
where energies and momenta are measured in the center-of-mass frame.  The yields 
were determined using a binned maximum-likelihood fit to the $\pipi$ missing mass spectrum for all
$\pipi$ pairs in selected events.  The fit function utilizes a background function composed of 
a simple polynomial to account for the combinatorial $\pipi$ background,  and a threshold function 
representing the onset of the inclusive $K_S$ threshold at $M_\text{miss}$ values of approximately 
$M(\upsv)-M(K_S)$.  To account for the signal due to $\pipi$ transitions from $\upsv$, 
a reversed Crystal Ball function (a normal Crystal Ball function\cite{cbfunction} with the tail on the high, rather than the low mass, side
of the curve) was used.  The shape parameters for the signal function
were obtained from a data sample of fully reconstructed transitions $\upsv\to\pipi\upsns$ (n=1,2,3) 
with $\upsns\to\mumu$.  This method not only provided a data-driven shape for the inclusive $\pipi$ signal, 
but also provided a check on the mass scale obtained using the $\pipi$ missing mass.

\begin{figure}[tb]
\centering
\includegraphics[width=0.8\textwidth]{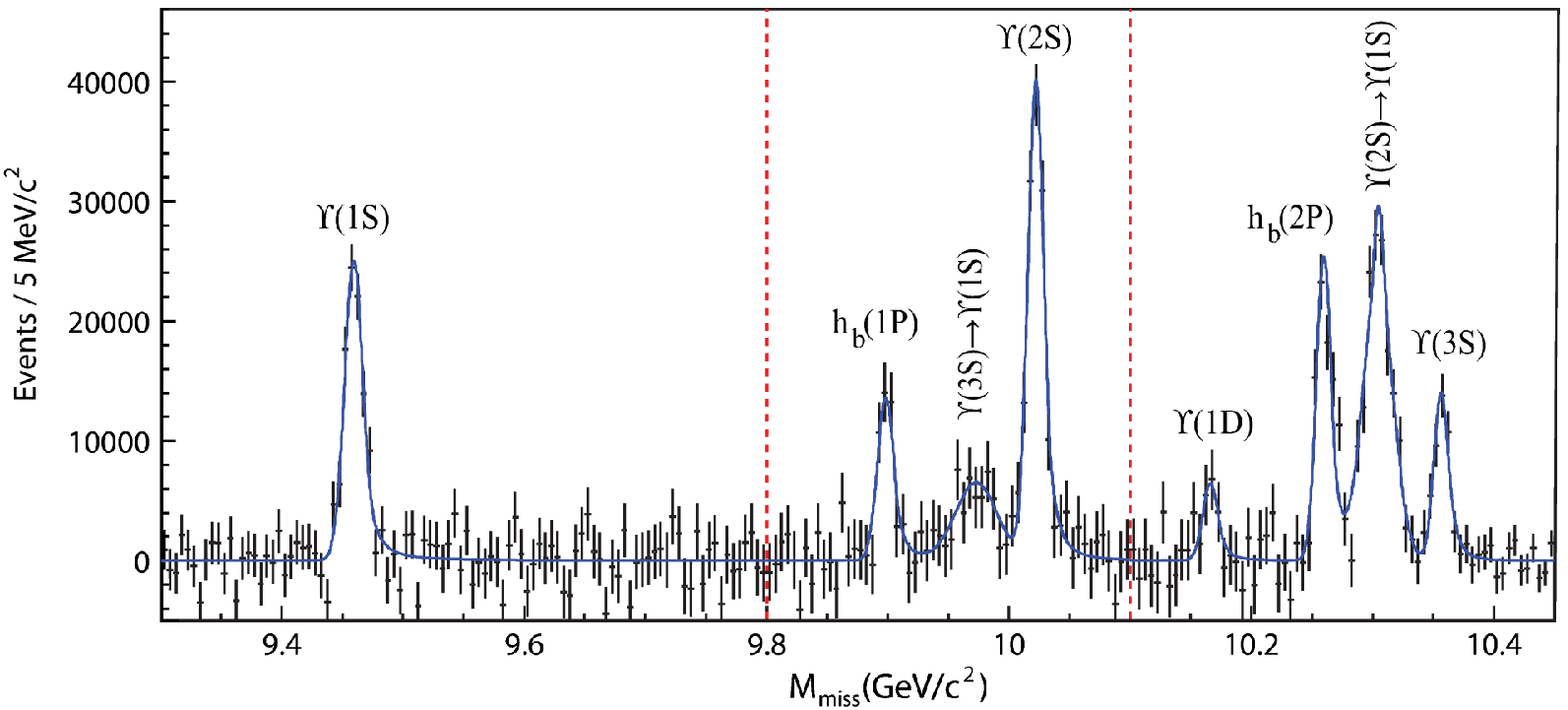}
\vspace*{8pt}
\caption{The spectrum of missing mass, $M_\text{miss}$, used by the Belle Collaboration
to search for $\upsv\to\pipi\hbn$ decays, shown after background subtraction.  The functional form of the fitted curve
is described in the text.  The peaks in the spectrum arise from direct transtions
$\upsv\to\pipi\Upsilon(1S,2S,3S)$, $\upsv\to\pipi\hbn$ and $\upsv\to\pipi\Upsilon(1D)$.
There are also peaks which are displaced from their expected location at $M(\upsi) = 9.46 \gevm$
due to cascade transitions $\upsv\to X + \Upsilon(3S,2S);
\Upsilon(3S,2S)\to\pipi\upsi$, wherein we observe only the lower transition $\pipi$ pair.  These peaks are labelled
$\upsiii\to\upsi$ and $\upsii\to\upsi$ in the spectrum.\label{fig:hb}
Reprinted figure with permission from I.~Adachi {\it et al.}  [Belle Collaboration],
Phys. Rev. Lett. 108, 032001, 2012. Copyright (2012) by the American Physical Society.}
\end{figure}

In the fit, signals corresponding to $\pipi$ transitions to 
all three $\upsns$, the two lower $\hbn$ states and $\Upsilon(1D)$ were used, as well
as functions corresponding to transitions $\upsiii\to\pipi\upsi$ and $\upsii\to\pipi\upsi$ 
in which the $\upsiii$ and $\upsii$ were produced inclusively in unobserved transitions 
from $\upsv$.
The final $\pipi$ missing mass spectrum is shown in Fig.~\ref{fig:hb}, with the fitted background subtracted, and 
with the fitted signal functions overlaid.  The significances
of the $\hb$ and $\hbp$ signals, with systematic uncertainties
accounted for, are $5.5\sigma$ and $11.2\sigma$, respectively.
These measurements represent the first observation of the singlet-P states of bottomonium.
Previously, there was only weak evidence for $\hb$ presented by \babar,
who sought it in the transition $\upsiii \to \piz\hb \to \piz \gamma \etab$.\cite{Lees11}

One important item to note is that, while the $\hbn$
search was prompted in part by the observation of anomalously large rates for $\pipi$ transitions
from $\upsv$, the rates for production of $\hbn$ obtained by this analysis
are also unexpectedly high.  The ratios $R\equiv\frac{\sigma(\Upsilon(\rm{5S}) \to \hbn\pipi)}{\sigma(\Upsilon(\rm{5S}) \to \Upsilon(\rm{2S}) \pipi)}$ were determined to be
$R=0.45\pm0.08\stat^{+0.07}_{-0.12}\syst$ for the $\hb$ and
$R=0.77\pm0.08\stat^{+0.22}_{-0.17}\syst$ for the $\hbp$.  Hence the same non-suppression of the
spin flip transition from $\upsv$ to $\hbn$ is observed as was observed in the charmonium case reported by CLEO.

The investigation of the reasons for this non-suppression ultimately led to the discovery of the charged $Z_b$ states,
which will be described later in Sec.~\ref{sec:chargedz}.  Nearly all the $\pipi$ transitions to the $\hbn$ states occur through
the $Z_b$ as an intermediate state, and this observation enabled a substantial decrease in the combinatorial background
(by a factor of 5 for the $\hb$ and 1.6 for the $\hbp$) by
requiring the observation of a $Z_b$ in the decay chain $\upsv\to\pi Z_b; Z_b\to\pi\hbn$.
Such background reduction enabled both, better $\hbn$ selection and more precise mass measurements with the result
$M(h_b(1P))  = (9899.1\pm 0.4 \pm 1.0)~{\rm MeV}/c^2$ and $M(h_b(2P)) = (10259.8 \pm 0.5 \pm 1.1)~{\rm MeV}/c^2$.\cite{Mizuk12}

\subsubsection{Observation of radiative transitions $\hbn\to\gamma\etabm$\label{sec:singlet-s}}

\begin{figure}[tb]
\centering
\includegraphics[width=0.9\textwidth]{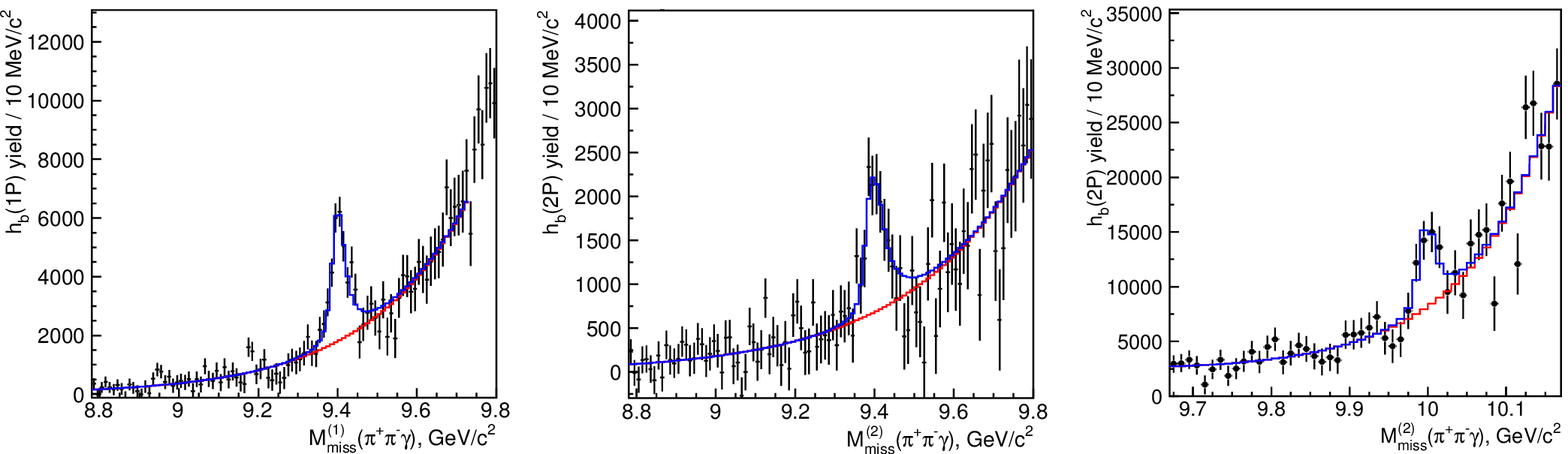}
\vspace*{8pt}
\caption{The spectrum of the variable $\mmissn(\pipi\gamma)$, used by Belle 
to search for $\hbn\to\gamma\etabm$ decays.\protect\cite{Mizuk12} A fit was performed to these data,
where the signal was modelled as a Crystal Ball line shape\protect\cite{cbfunction} convoluted with a Breit-Wigner function,
and the background was modelled as exponential polynomial. The peaks in the spectrum arise from $\hb\to\gamma\etab$ (left),  $\hbp\to\gamma\etab$ (middle) and $\hbp\to\gamma\etabp$ (right).\label{fig:ppg}
Figure courtesy of the Belle Collaboration.}
\end{figure}

Subsequent to the observation of the singlet-P states, Belle launched a
study of the expected principal decay modes of the singlet-P states, namely,
the E1 transitions $\hb \to \gamma \etab$, $\hbp \to \gamma \etab$, and $\hbp \to \gamma
\etabp$.\cite{Mizuk12} The branching fractions for these three transitions were
predicted by Godfrey and Rosner to be $41.4\%$, $12.5\%$ and $19.3\%$, respectively.\cite{godros}
The search involved selection of events broadly consistent with the production via $\pipi$ transition of either $\hbn$ state,
consistent with the intermediate production of a $Z_b$ state, and the observation of a photon.
A two-dimensional method was then employed, in which the $\hbn$ yield was determined in bins of
the variable $M_{\rm miss}^{(m)} (\pipi\gamma) = M_{\rm miss}(\pipi\gamma) - M_{\rm miss}(\pipi) + M(\hbn)$.
The $\etabm$ yield was obtained by binned maximum-likelihood fits to the variable $\mmissn(\pipi\gamma)$.
The distributions of this variable for events corresponding to each transition are shown in Fig.~\ref{fig:ppg}.

\begin{table}[tb]
\centering
\caption{E1 branching fractions for bottomonium singlet-P states.  Errors cited are statistical and systematic, respectively.}
\begin{tabular}{ccc}
\hline \hline
Transition      & Branching Fraction\cite{Mizuk12} & Prediction\cite{godros}\\
\hline
$\hb\to\gamma\etab$ & $(49.2\pm 5.7 \asy{5.6}{3.3})\%$ & $41.4\%$ \\
$\hbp\to\gamma\etab$ & $(22.3\pm 3.8\asy{3.1}{3.3})\%$ & $12.5\%$\\
$\hbp\to\gamma\etabp$ & $(47.5\pm 10.5\asy{6.8}{7.7})\%$ & $19.3\%$\\
\hline \hline
\end{tabular}\label{ta2} 
\end{table}

These investigations yielded the first observation of the radial excitation of $\etab$, namely $\etabp$, with
$M(\etabp) = (9999.0 \pm 3.5\stat^{+2.8}_{-1.9}\syst)$
MeV/$c^2$, and measurements of the branching fractions for $\hbn\to\gamma\etabm$ ($m^\prime = 1,2$) (see Table~\ref{ta2}).  The resulting
hyperfine splitting in the 2S level, of $\Delta M_{\rm HF}(2S) = 24.3 \asy{4.0}{4.5} {\rm MeV}/c^2$ was  found to be in agreement with theoretical expectations,\cite{Meinel10,Dowdall12} while the branching fractions
were, in general, larger than the predicted values.\cite{godros}  A 90\% confidence level upper limit for the width of $\etabp$ was set at $24$ MeV$/c^2$.
In addition, the combined samples of events in which the $\etab$ was observed enabled Belle to make the
world's most precise measurement of the $\etab$ mass, $M(\etab) = (9402.4\pm 1.5\stat \pm 1.8\syst)$ MeV/$c^2$, and to measure its
width for the first time: $\Gamma(\etab) = 11\asy{6}{4}~{\rm MeV}/c^2$.
The 1S hyperfine splitting of $\Delta M_{\rm HF}(1S) = 57.9\pm 2.3$ MeV$/c^2$
that corresponds to the $\etab$ mass measurement is in much better agreement with theoretical expectations than are previous
measurements.\cite{Meinel10,Dowdall12}

\subsubsection{Observation of $\pipi$ transitions to $D$-wave states\label{sec:y1d}}

Among the peaks observed in the $\pipi$ missing mass distribution for the inclusive $\upsv\to\pipi + X$ analysis\cite{Adachi12} (see Fig.~\ref{fig:hb}) is a peak that corresponds to $\pipi$ transitions to the $\upsid$ states at 10.16 GeV.  In the inclusive
analysis, the significance of the peak was insufficient (at only  $2.6\sigma$) to claim observation.  Belle undertook a fully exclusive analysis in order to establish observation of the $\upsid$ state, reconstructing the full decay chain $\upsv\to\pipi\upsid$; $\upsid\goesto\gamma\chibj$; $\chibj\goesto\gamma\upsi$; $\upsi\goesto\mumu$, and establishing a signal for $\upsv\to\pipi\upsid$ at the $9\sigma$ level of significance.
Interestingly, the observed yield indicates 
a partial width for the $\upsv\to\pipi\upsid$ of $\sim 60$ keV$/c^2$, which is much larger than expected.\cite{Mizuk13}

\subsubsection{Observation of $\eta$ transitions to $\Upsilon(1S,2S)$\label{sec:etatrans}}

Transitions between vector meson states via the emission of an $\eta$ meson are of interest historically in part because of the observation of a larger than expected branching fraction for $\psi(2S)\to\eta\jpsi$, $(3.3\pm 0.5)\%$,
compared to that for $\psi(2S)\to\pipi\jpsi$, $(34.0\pm 0.4)\%.$\cite{pdg}
The QCD multipole expansion model\cite{Gottfried78,Voloshin79,Yan80}
allows one to classify hadronic transitions between heavy quarkonium states 
as arising from the emission and subsequent hadronization of a pair of gluons
that are emitted in various combinations of chromo-electric or chromo-magnetic multipoles.  
The simplest such transitions, $\pipi$ transitions, occur due to the emission of, in lowest order,
a pair of chromo-electric dipole (E1) gluons.  $\eta$ transitions require a higher order combination, 
an E1 gluon and a chromo-magnetic quadrupole (M2) gluon.  Hence, in the transition between any two vector states,
the rate for the $\eta$ transition ought to be substantially suppressed relative to that for the
$\pipi$ transition between the same states.\cite{Kuang,Voloshin}

A recent Belle observation of the transition $\upsii\to\eta\Upsilon(1S)$ was consistent with the expectation of suppression,
in which the ratio of rates for $\eta$ to that for the corresponding $\pipi$ transition of
$(1.99\pm 0.14(\text{stat}) \asy{0.12}{0.08}(\text{syst}))\times 10^{-3}$ was measured.\cite{belle2seta}  \babar, however, observed an
unexpectedly high rate for the $\upsiv\to\eta\upsi$ transition, measuring an $\eta$ to $\pipi$ rate ratio of
$2.41 \pm 0.40(\text{stat}) \pm 0.21(\text{syst})$.\cite{Aubert08a}  This result indicated a possible breakdown of the QCD multipole expansion model for transitions
from states lying above open-flavor threshold, and motivated a search for $\eta$ transitions from the $\upsv$ by Belle.

The Belle analysis of the transitions $\upsv\to\eta(\upsi,\upsii)$ involved full reconstruction of the entire decay chain, with
$\eta\to\gamma\gamma$ and $(\upsi,\upsii)\to\mumu$.
Again, the ratio of rates for the $\eta$ transition was measured relative
to that for the corresponding $\pipi$ transitions, and it was found that the $\eta$ transitions from $\upsv$ are also not substantially
suppressed relative to the $\pipi$ transitions.  
Ratios of $0.16\pm 0.04(\text{stat}) \pm 0.02(\text{syst})$ and $0.48\pm 0.05(\text{stat}) \pm 0.09(\text{syst})$ for the 
transitions to $\upsi$ and to $\upsii$, respectively, were measured.\cite{Mizuk13}  Attempts to explain the lack of suppression of these
$\eta$ transitions incorporate the possibility of either a tetraquark resonant substructure in the parent wave function, or rescattering through
$B\Bbar$ pairs.\cite{volo_eta}

\subsection{Spectroscopy of unconventional bottomonium states}

Quantum Chromodynamics does not limit hadronic structures to configurations
involving three quarks (baryons) and those involving a quark and an antiquark (mesons).  It has in fact
been something of a surprise that clear examples of tetraquarks ($q q \bar q \bar q$),
pentaquarks ($qqqq \bar q$) or meson molecules $(q\bar q)(q'\bar q')$ have not been
unambiguously identified. One of the most important results of
the study of the Belle $\upsv$ data sample is the discovery of a number of interesting
states which are unambiguous examples of such unconventional structures.

\subsubsection{$Z_b^{\pm}(10610)$ and $Z_b^{\pm}(10650)$\label{sec:chargedz}}

If the dipion transitions from $\upsv$ to lower bottomonium states, both
$\upsv\to\pipi\upsns$ and $\upsv\to\pipi\hbn$ proceeded
by the emission of two gluons from the initial $b\bar b$ state, then the production of
vector $\upsns$ states would dominate, since the transition to $\hbn$
requires a $b$-quark spin flip and is expected to be heavily suppressed.
This is not what was observed, however -- the ratios of partial widths for $\pipi\hbn$ to those of
$\pipi\upsns$ are of order one. If instead of being directly produced in the transition from
$\upsv$, the $\pi^{\pm}$ were sequentially produced in a cascade of decays,
the expected suppression of $\hbn$ production would not necessarily occur.
It is this fact that led to search for an explanation in the resonant substructure of the
$\upsns\pipi$ and $\hbn\pipi$ final states.

\begin{figure}[tb]
\centering
\includegraphics[width=0.9\textwidth]{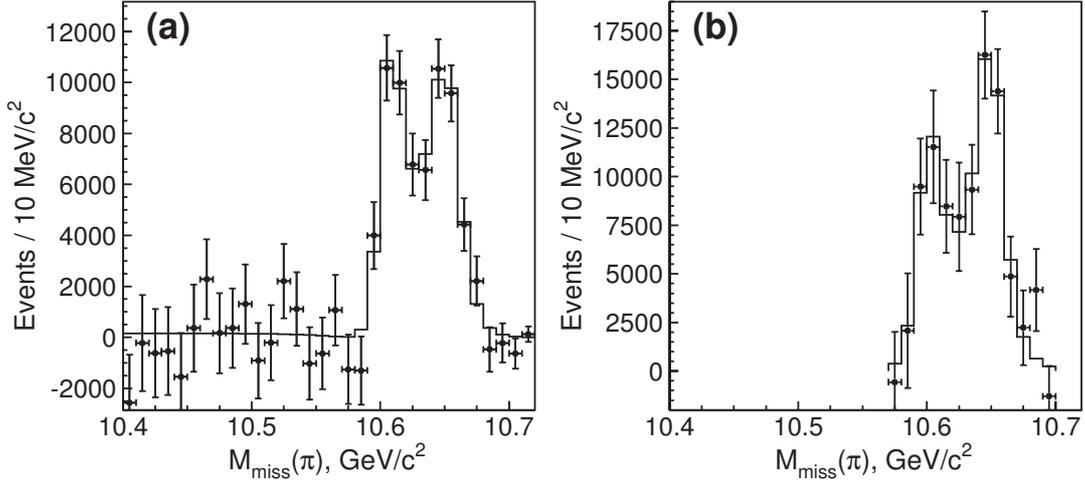}
\vspace*{8pt}
\caption{The $M_{\rm miss}(\pi)$ spectra for events consistent with the decay chain
$\upsv\to\pipi\hb$ (left) and $\upsv\to\pipi\hbp$ (right). These missing masses are equivalent to the
invariant mass of $\hb\pi$ and $\hbp\pi$, respectively. \label{fig:zb}
Reprinted figure with permission from A.~Bondar {\it et al.}  [Belle Collaboration],
Phys. Rev.  Lett. 108, 122001, 2012. Copyright (2012) by the American Physical Society.}
\end{figure}

A study of the invariant mass of $\pi^{\pm}\upsns$ and $\pi^{\pm}\hb$ revealed
resonances having masses between $\upsiv$ and $\upsv$.  Signals of each of these states, which
are electrically charged, were clearly observed in each of five different decay channels,
$\Upsilon({\rm nS})\pipm$ (n = 1, 2, 3) and $\hbn\pipm$ (m = 1, 2).
These resonances are denoted $Z_b^{\pm}(10610)$ and $Z_b^{\pm}(10650)$.\cite{Bondar12} In Figure~\ref{fig:zb} are
shown the single-pion missing masses $M_{\rm miss}(\pi)$ for events consistent with $\pipi$ transitions
to $\hb$ and $\hbp$.

Because of their large masses, these states necessarily have a bottom
quark and antiquark as constituents -- but because they are charged, they must also include
another pair of quarks, and therefore they are unambiguously unconventional in their quark structure.
Whether they are simply a tetraquark state, e.g. $|Z_b^+\rangle \equiv |b\bar b u\bar d\rangle$, or a
molecular state remains to be seen.   One interesting point worth considering is the relative
proximity of their masses to the $B\Bstar$ and $\Bstar\Bstar$ thresholds,
which lends credence to the possibility that they are molecular states containing a $B^{(*)}\bar B^*$ pair.
Such a description predicts equal total widths (as
observed) and large branching fractions to the appropriate $B^{(*)}\bar B^*$ final state (also as observed, as discussed below).
The properties of these states are summarized in Table~\ref{ta3}.  Both
have isospin $I=1$, positive G-parity, and are determined to have spin-parity
$J^P = 1^+$ by angular analysis of their production and decay kinematics.

\begin{table}[tb]
\centering
\caption{Properties of multiquark bottomonium-like states.}
\begin{tabular}{cccc}
\hline \hline
State      & Mass     & Width & Reference\\
       & (MeV/$c^2$)  & (MeV/$c^2$) & \\
\hline
$Z_b^{\pm}(10610)$ & $10607.2\pm2.0$ & $18.4\pm2.4$ & \cite{Bondar12}\\
$Z_b^{\pm}(10650)$ & $10652.2\pm1.5$ & $11.5\pm2.2$ & \cite{Bondar12}\\
$Z_b^{0}(10610)$ & $10609 \pm 4 \pm 4$ & $-$ & \cite{Krokovny:2013mgx}  \\
\hline \hline
\end{tabular}
\label{ta3}
\end{table}

In an attempt to further elucidate the nature of these states, Belle investigated the resonant substructure of
three-body final states of $[B^{(*)}B^{(*)}]^{\pm}\pi^{\mp}$.\cite{Adachi12b}  In this analysis, one of the
daughter $B^{(*)}$ mesons was fully reconstructed, while the other was inferred using the missing mass of
the reconstructed $B^{(*)}\pi$ system. It was observed that the
$Z_b^{\pm}(10610)$ decays with a branching fraction of $(86.0\pm 3.6)\%$ to $BB^*$,
and that the $Z_b^{\pm}(10650)$ decays with a branching fraction of $(73.4\pm 7.0)\%$ to
$B^*B^*$. These branching fractions are calculated under the assumption that the $Z_b^{\pm}$ decays
solely to $B^{(*)}B^*$, $\pipm\upsns$ and $\pipm\hbn$.
This result does not represent definitive proof of the molecular nature of the $Z_b$ states,
but is strong evidence in its favor.

\subsubsection{$Z_b^0(10610)$\label{sec:neutralz}}

Naturally, once Belle had identified charged $Z_b$ states, one might additionally expect the existence
of neutral isospin partners.  In order to search for them,  the analogous $\pizpiz$
transitions $\upsv\to\pizpiz\upsns$ were investigated and the final-state Dalitz plot was treated in a manner similar
to that in the previously-described charged dipion study.
The $Z_b^0(10610)$ state was observed in the single $\piz$ missing mass with a significance at the 
$6.5\sigma$ level, performing a combined fit to the $\Upsilon(2S) \pi^0 \pi^0$ and $\Upsilon(3S) \pi^0 \pi^0$
samples.\cite{Krokovny:2013mgx} The measured mass of the state, $(10609 \pm 4 \stat\pm 4\syst)$ MeV$/c^2$, suggests that it is the isospin partner of the charged $Z_b^\pm(10610)$. In addition, there is slight evidence for an isospin partner for the higher
mass charged $Z_b$, but the statistical significance of about $2\sigma$ is insufficient to claim observation.

\subsubsection{$Y_b(10890)$\label{sec:yb}}

In 2008, as noted above in Sec.~\ref{sec:pipirates}, Belle observed rates for transitions
$\upsv\goesto\pipi\upsot$ that were much larger than theoretical expectations.\cite{Chen08}
To investigate this anomaly, the $\pipi\upsot$ cross section was measured for center-of-mass
energies in the range between 10.83 $\gev$ and 11.02 $\gev$ to search for potential states in additition to the conventional $\upsv$.
A peak was found in the cross section $\sigma(e^+e^- \to \upsns \pi^+ \pi^-)$ (n = 1, 2, 3) at $(10888^{+2.7}_{-2.6}\stat\pm 1.2\syst)$ MeV$/c^2$ with a width of $(30.7^{+8.3}_{-7.0} \stat\pm 3.1\syst)$ MeV$/c^2$.\cite{Chen:2008xia}
The peak values and widths observed for transitions to each of the three lower $\upsns$ states were mutually consistent.
The fact that this value is displaced from the peak in the $b\bar b$ cross section,\cite{Aubert09} led to the
inference of a possible exotic state $Y_b$ that is nearly degenerate with $\upsv$.\cite{Liu12,Ali13}

\section{Outlook}

The large $\Upsilon(5S)$ data sample collected by the Belle experiment made possible numerous
measurements of $B_s$ decays and provided the unexpected possibility to study conventional and exotic
bottomonium states. Many of these decays and bottomonium states were observed for the first time.
Their observation calls for further high precision studies and brings up new questions, that
require a considerably larger sample of $\Upsilon(5S)$ data. Such a sample will be collected by
Belle II at SuperKEKB, the upgrade to the Belle detector. The accelerator and the detector will be online from 2015.
The target instantaneous luminosity will be $8 \times 10^{35}$ cm$^{-2}/{\rm s}^{-1}$, 40 times that of KEKB.
Assuming the same data collection ratio of $\Upsilon(5S)/\Upsilon(4S)$ as that at Belle, we can anticipate an
$\Upsilon(5S)$ sample of as much as 5 ab$^{-1}$ of data.

A 5 ab$^{-1}$ $\Upsilon(5S)$ sample contains approximately 300 million $B_s \bar B_s$ pairs and will allow for
comprehensive studies of the decay
rates of the $B_s$ with a completeness and accuracy comparable to that currently available for $B^0$ and $B^+$ mesons, thereby
improving our understanding of $B$ physics. Comparative studies of $B^0$ and $B_s$ mesons will help
to reduce the theoretical uncertainties related to quantities sensitive to new physics.
Moreover, $B_s$ physics provides additional opportunities to probe new physics effects in $b \to s$ transitions.
The most notable improvement for $B_s$ measurements is the possibility to exploit tagging techniques that have been so
successfully applied at $\Upsilon(4S)$, allowing for high purity measurements of decays with neutrals and missing energy,
and the reduction of uncertainties due to $f_s$ to only a few percent. The golden new physics search modes of Belle II will be the
flavor-changing neutral-current transitions suppressed in the SM:
$B_s \to \tau \tau$ (${\cal B}=8.9\times10^{-7}$)\cite{Bobeth:2011st,Dighe:2012df}, $B_s\to \nu \bar\nu(\gamma)$
(${\cal B}=7.5\times10^{-8}$)\cite{Badin:2010uh,Dighe:2012df} and $B_s \to \gamma\gamma$
(${\cal B}=(0.7^{+2.5}_{-0.4})\times10^{-6}$)\cite{Reina:1997my,Gemintern:2004bw}, which complement the high
profile searches of $B_s \to \mu\mu$ at the LHC.
Various outstanding problems in semileptonic $B$ decays can be well complemented by precise measurements in $B_s$ decays,  where the
heavier strange spectator quark leads to smaller theoretical uncertainties.\cite{Urquijo:2013npa}
Precise absolute branching fractions of Cabibbo-favored transitions, {\it e.g.} $B_s\to D_s^{(*)}h$,
will test predictions of QCD in $B_s$ decays and SU(3) symmetry.\cite{Fleischer:2010ca}
With minimal trigger bias, Belle II can perform complete surveys of the full range of $B_s$ decay modes to complement programs
at hadron machines. Despite the $B_s$ having an oscillation frequency too fast for measurements of time-dependent $CP$ violation
at Belle II, the experiment can still provide unique information on the weak mixing amplitude in $B_s$ decays with neutral final states,
as well as studies of time independent $CP$ violation.

Belle II will also play a leading role in the study of bottomonium and other hadron physics at the $\Upsilon(5S)$. These studies
are highly sensitive to physics triggers, and hence are very challenging to perform at hadron colliders.
With numerous discoveries and many unsolved puzzles, spectroscopy will move from an era of first observations to
precision measurements, to clarify quantum numbers and states. 
The charged $Z_b^\pm$ states and the neutral $Z_b^0(10610)$, discovered by Belle, represent candidates for $B$-meson molecules.
Whether these are bound states should be determined by proving the existence of a second neutral state, $Z_b^0(10650)$,
and studies of radiative transitions, $Z_b \to B^{(*)} \bar{B} \gamma$.\cite{Drutskoy:2012gt} There is also interest in the search
for possible sibling states, $W_{b{\rm J}}^{(\prime)}$ (J = 0, 1, 2), decaying to $\chi_b$ or $\eta_b$, and light
hadrons.\cite{Bondar:2011ev} Future studies based on the missing mass method will profit from the considerably larger data sample.
The $Y_b(10890)$, a candidate for a tetraquark state, attracted strong interest in the theory
community. The measurement of $B^{(*)}\bar{B}^{(*)}$ and $B_s^{(*)}\bar{B}_s^{(*)}$ production rates at the $\Upsilon(5S)$ energy
inconsistent with expectations from SU(3) symmetry is another hint for the existence of this exotic state and needs
further investigation.\cite{Drutskoy:2012gt} There are also candidates for tetraquark and molecule states in the
charm sector, but currently no theory provides a consistent picture of the whole spectrum.\cite{Hambrock:2013tpa} The
large $\Upsilon(5S)$ sample that will become available at Belle II will help to clarify the issues currently under discussion
and will allow for many other physics studies, uniquely feasible at an $e^+e^-$ collider.\cite{Drutskoy:2012gt}

\section*{Acknowledgments}

We would like to thank Phillip Urquijo and the Belle collaboration for useful comments
to this paper and their support. The work of C. Oswald was supported by a
doctoral scholarship of the University of Bonn. The work of T. K. Pedlar was supported in part
by the National Science Foundation through Grant No.\ PHY-1205843.


\begin{thebibliography}{0}
\bibitem{BsCrosss}
The estimation of the $B_s$ production at the $\Upsilon(5S)$ resonance is described below.
\end{thebibliography}
\end{document}